\documentclass[english,a4paper,12pt]{article}

\usepackage{babel}
\usepackage[T1]{fontenc}
\usepackage{amssymb}
\usepackage[dvips]{graphicx}

\usepackage{hyperref}

\usepackage{soul}

\newtheorem{tm}{Theorem}
\newtheorem{df}{Definition}
\newtheorem{lem}[tm]{Lemma}
\newtheorem{cor}[tm]{Corollary}

\newcommand{\refeq}[1]{\textup{(}\ref{eq:#1}\textup{)}}

\usepackage{xcolor}
\definecolor{cobalt}{rgb}{0.0, 0.28, 0.67}

\newcommand{\takeout}[1]{}

\title{A continuum of incomplete intermediate logics\thanks{This article is
based on a paper delivered at the 4th International Tbilisi
Symposium on Language, Logic and Computation (September 2001).} \\
(corrected version)}
\author{Tadeusz Litak\\ \st{Department of Logic, Jagiellonian
University}\\ \st{Grodzka 52, 31-044 Cracow}\\ 
\st{tlt@konto.pl}}
\date{}

\newsavebox{\storey}
\savebox{\storey}(0,0)[bl]{\put(0,0){\line(1,1){20}}
\put(20,0){\line(-1,1){20}} \multiput(0,20)(20,0){2}{\circle*{1}}
\multiput(0,20)(20,0){2}{\line(0,-1){10}}}

\newsavebox{\fshup}
\savebox{\fshup}(0,0)[bl]{\put(-10,0){\line(-3,1){30}}
\put(-10,0){\line(1,1){10}}\put(-10,0){\circle*{1.5}}
\put(-20,0){\circle*{1.5}}
\put(-40,10){\line(2,1){20}}\put(-40,10){\circle*{1.5}}
\put(-20,20){\line(0,-1){20}}\put(-20,20){\circle*{1.5}}}

\newsavebox{\fshdown}
\savebox{\fshdown}(0,0)[bl]{\put(0,10){\line(2,1){20}}
\put(20,20){\line(-4,-1){80}}\put(20,20){\circle*{1.5}}}

\newsavebox{\fsh}
\savebox{\fsh}(0,0)[bl]{\put(0,0){\usebox{\fshup}}
\put(0,0){\usebox{\fshdown}}}

\newcommand{\tmlei}[1]{\textbf{TML2018: #1}}

\begin{document}

\maketitle

{ \scriptsize
\textbf{Note 2018:} \textcolor{cobalt}{This paper was originally published in Reports on Mathematical Logic 36, 2002, pp. 131--141. I have recently noted that the proof of one of theorems in it was incorrect; it was also independently discovered by Guillaume Massas (UC Irvine). This does not concern the main result claimed in the title (Theorem \ref{t:main}), which seems unassailable, but rather my attempt to present the proof  of  Theorem \ref{t:incompl}, essentially due to Shehtman, without, as I say below, ``a superfluous use of transfinite induction'' (i.e., differing with the original paper \cite{shehtman} and  my own Master's Thesis). My version of proof is fixable; I would like to thank Guillaume for coming up with the idea. Hopefully, some of his work towards generalizing such results will be published soon. Let us also note that Valentin Shehtman himself points out that the proof in the 1980 paper \cite{sh80} or in his more recent Habilitation Thesis has already been simplified compared with the one in the original reference \cite{shehtman}.  
Apart from this crucial fix (and adjusting one reference), I left this earliest paper of mine unchanged on principle, even though I was tempted to polish up---at the very least---its style, narration and English.} 
}

\begin{abstract}
Although in 1977 V.B. Shehtman constructed the first Kripke
incomplete intermediate logic, no-one in the known literature has
completed his work by constructing a continuum of such logics.
After a substantial reminder on how an incomplete logic can be
obtained, I will construct a sequence of frames similar to those
used by Jankov and Fine. None of these frames can be reduced by a
p-morphism to another; at the same time, there are no p-morphisms
from generated subframes of the Fine frame onto any frame from the
considered sequence. All of the frames satisfy all of Shehtman's
axioms. Therefore, by using the characteristic formulas of the
frames from the sequence it is possible to obtain the desired
conclusion.
\end{abstract}

In the 1970's, a number of important, deep and technically
complicated results concerning relational semantics for modal
logics was obtained by such authors as S. Thomason, K. Fine, M.S.
Gerson, R.I. Goldblatt, J. F. A. K. van Benthem and W. Blok; it
was the Golden Age of the subject, see \cite{bull82},
\cite{bull83} and \cite{chagrov} for references and summaries of
the most important works. The main goal of my paper is to draw
attention to the fact that many important results lack
superintuitionistic analogues, although the task of transferring
them is highly nontrivial.

This gap may be partially due to the fact that Kripke semantics
never became as popular in the realm of intermediate logics as
they are in the realm of modal logics, which are more suitable and
flexible tools to deal with frames. There were fewer experts
working on relational semantics for intuitionistic logics. In
1977, one of the most distinguished persons in the field, V. B.
Shehtman, constructed the first Kripke incomplete intermediate
propositional logic. His construction was based mainly on a frame
from~\cite{fine74b}, but he very ingeniously used a formula
introduced in~\cite{gabbay}. Nevertheless, he did not follow
Fine's suggestion that it seems to be possible to construct a
continuum of incomplete logics. Such a continuum of $S4$ logics
was presented in~\cite{rybakov} in the same year as Shehtman's
construction; it is known, however, that the incompleteness of a
modal logic does not imply the incompleteness of its
intuitionistic equivalent. In~\cite{ono} one may find the claim
that there exists a continuum of incomplete predicate
superintuitionistic logics. Unfortunately, this claim is given
without proof; besides, it is far easier to construct an
incomplete predicate superintuitionistic logic than to construct
an incomplete propositional superintuitionistic  logic. It is
truly surprising but up to this day no-one has presented a proof
that there exists a continuum of such logics. I shall attempt to
fill in this gap.

In this paper I shall try to conform to the standard definitions
and symbols which may be found, for example, in a monograph by
Chagrov \& Zakharyaschev~\cite{chagrov}. Nevertheless, for the
sake of convenience, let me remind the most standard ones. Unless
otherwise stated, by a logic I shall mean a superintuitionistic
(intermediate) logic.

\begin{df}
\emph{A (Kripke) structure/frame} consists of a set and a relation
of partial order ${\cal F}~=~\langle W, \leqslant\rangle$.
\end{df}

\begin{df}
\emph{A substructure/subframe} of a structure ${\cal F}~=~\langle
W, \leqslant \rangle$ is a frame ${\cal G}~=~\langle V,
\leqslant_1 \rangle$ where $V~\subseteq~W$ and
$\leqslant_1~=~V^2~\cap~\leqslant$.
\end{df}

\begin{df}
\emph{A (Kripke) model} is an ordered pair ${\cal M}~=~\langle
{\cal F}, {\cal B} \rangle$ consisting of a frame ${\cal
F}~=~\langle W, \leqslant \rangle$ and a function ${\cal B}$ from
the set of propositional variables to the set of upward closed
subsets of~$W$. Valuation is extended to all formulas in the usual
way.
\end{df}

I would like now to introduce two technical notions, weaker than
\emph{finite approximability} (\emph{finite model property}) and
stronger than \emph{completeness}

\begin{df}
A logic is \emph{fa-approximable} iff the set of its theorems
coincides with the set of all formulas which are true in some
class of rooted frames with no infinite antichains.
\end{df}

\begin{df}
A logic is \emph{ac-approximable} iff the set of its theorems
coincides with the set of all formulas true in some class of
frames with no infinite ascending chains --- Chagrov \&
Zakharayaschev call such orders \emph{Noetherian}.
\end{df}

Professor A. Wro\'{n}ski has suggested that fa-approximability
implies ac-approximability. This would give rise to the following
picture:

\begin{center}
finite approximability~ $\Rightarrow$~ fa-approximability~
$\Rightarrow$~ ac-approximability~ $\Rightarrow$~ completeness.
\end{center}

In my paper, I shall prove that there exists a continuum of
propositional logics even outside the broadest class, i.e. the
class of all complete logics. Nevertheless, first let me describe
how an incomplete logic can be obtained --- it is an easy
generalization of Shehtman's method \cite{shehtman}.

\begin{tm} \label{t:grz}
A logic $L$ lacks ac-approximability iff its
modal companion above {\bf Grz} $\tau\!L$ is incomplete.
\end{tm}

{\sc Proof.} It is enough to recall that {\bf Grz} is complete
with respect to all partial orders without infinite ascending
chains. \hfill $\dashv$ \vspace{0.2 cm}

\begin{tm} \label{t:ac}
If there exists a rule of the form
\[ \frac{(\psi~\vee~(\psi~\rightarrow~e(\chi)))~\rightarrow~\chi}
{\chi} \] ($e$ is any uniform substitution) which is not
admissible in some intermediate logic, then this logic lacks
ac-approximability and thus lacks the finite model property.
\end{tm}

{\sc Proof.} (sketch) In any family of frames adequate for the
logic (if there exists such) there must be a frame validating
\[(\psi~\vee~(\psi~\rightarrow~e(\chi)))~\rightarrow~\chi \] with
all substitutions (because the formula belongs to the logic) and
refuting $\chi$ under some valuation. It can be easily seen that
such a frame must contain an infinite ascending chain --- see
figure ~\ref{r:ac}. \hfill $\dashv$ \vspace{0.2 cm}

\setlength{\unitlength}{0.7mm}

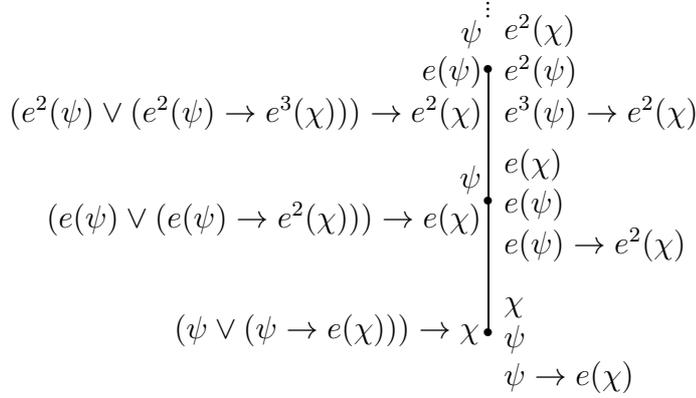
\begin{figure}[t]
\begin{center}
\begin{picture} (140,70)

    \multiput(70,10)(0,25){3}{\circle*{1.5}}
    \put(70,10){\line(0,1){50}}
    \multiput(70,70)(0,1.25){3}{\circle*{0.3}}

    \put(11,9){\shortstack[r]{$(\psi
\vee (\psi \rightarrow e(\chi))) \rightarrow \chi$}}
    \put(73,0){\shortstack[l]{$\chi$ \\ $\psi$ \\ $\psi \rightarrow
    e(\chi)$}}

    \put(-13,30){\shortstack[r]{$\psi$ \\ $(e(\psi)
\vee (e(\psi) \rightarrow e^2(\chi))) \rightarrow e(\chi)$}}
    \put(73,25){\shortstack[l]{$e(\chi)$ \\ $e(\psi)$ \\ $e(\psi) \rightarrow
    e^2(\chi)$}}

    \put(-20,50){\shortstack[r]{$\psi$ \\ $e(\psi)$ \\ $(e^2(\psi)
\vee (e^2(\psi) \rightarrow e^3(\chi))) \rightarrow e^2(\chi)$}}
    \put(73,50){\shortstack[l]{$e^2(\chi)$ \\ $e^2(\psi)$ \\ $e^3(\psi) \rightarrow
    e^2(\chi)$}}

\end{picture}
\caption{A model refuting $\chi$ but verifying $\psi~\vee~(\psi~
\rightarrow~e(\chi))~\rightarrow~\chi$ \label{r:ac}}
\end{center}
\end{figure}

\setlength{\unitlength}{0.8mm}

\begin{cor}
If an intermediate logic satisfies the assumptions of
theorem~\ref{t:ac}, then its companion above {\bf Grz} is
incomplete.
\end{cor}

{\sc Proof.} A consequence of theorems~\ref{t:grz}
and~\ref{t:ac}.\hfill $\dashv$ \vspace{0.2 cm}

In fact far more can be proved about such a logic --- see my
forthcoming paper~\cite{li}.

\begin{figure} [b]
\begin{center}
\begin{picture} (50, 50)

    \put(10,10){\line(1,1){30}}

    \multiput(10,10)(10,10){4}{\circle*{1.5}}
    \multiput(10,10)(10,10){4}{\line(-1,1){10}}
    \multiput(0,20)(10,10){4}{\circle*{1.5}}

    \multiput(45,45)(1.25,1.25){4}{\circle*{0.3}}

\end{picture}
\caption{An infinite comb \label{r:comb}}
\end{center}
\end{figure}
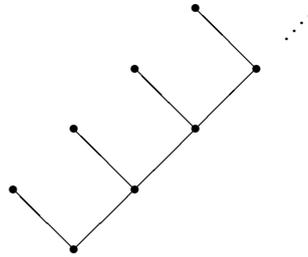

\begin{tm}
If there exists a rule of the form
\[ \frac{\begin{array}{ccc} (\psi~\vee~(\psi~\rightarrow~e(\chi)))~
\rightarrow~\chi \\  \psi~\leftrightarrow~\varsigma~\rightarrow~
\tau \\ \tau~\rightarrow~e(\tau) \end{array}}{\chi}
\] which is not admissible in a logic $L$, then in any
class of frames adequate for $L$ (if there exists any) there must
be a structure containing \emph{an infinite comb} or \emph{a
willow} (see fig.~\ref{r:comb}) as a substructure; thus, $L$ must
lack both ac-approximability and fa-approximability.
\end{tm}

\begin{figure} [t]
\begin{center}
\begin{picture} (50, 50)

    \put(10,10){\line(1,1){30}}

    \multiput(10,10)(10,10){4}{\circle*{1.5}}
    \multiput(10,10)(10,10){4}{\line(-1,1){15}}
    \multiput(-5,25)(10,10){4}{\circle*{1.5}}

    \multiput(45,45)(1.25,1.25){4}{\circle*{0.3}}

    \put(-9,24){$\varsigma$}
    \put(-11,34){$\psi~e(\varsigma)$}
    \put(-14,44){$\psi~e(\psi)~e^2(\varsigma)$}
    \put(-17,54){$\psi~e(\psi)~e^2(\psi)~e^3(\varsigma)$}

    \put(-3,24){$\tau$}
    \put(7,34){$e(\tau)$}
    \put(17,44){$e^2(\tau)$}
    \put(27,54){$e^3(\tau)$}

    \put(13,9){$\chi$}
    \put(23,19){$e(\chi)$}
    \put(33,29){$e^2(\chi)$}
    \put(43,39){$e^3(\chi)$}

    \put(14,19){$\psi$}
    \put(17,29){$e(\psi)$}
    \put(26,39){$e^2(\psi)$}

\end{picture}
\caption{A submodel of $\langle {\cal F}, {\cal V}\rangle$ whose
root refutes $\chi$. \label{f:comb}}
\end{center}
\end{figure}
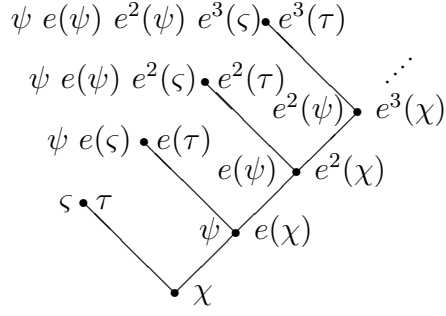

{\sc Proof.} Similar to the proof of theorem~\ref{t:ac} --- see
fig.~\ref{f:comb}. \hfill $\dashv$ \vspace{0.2 cm}

Let me recall the celebrated Gabbay-de Jongh axioms~\cite{gabbay}

\[ \mbox{\boldmath$bb_n$}~:=~\bigwedge\limits^n_{i~=~0}
((p_i~\rightarrow~\bigvee\limits_{j~\neq~i}p_j)~\rightarrow~
\bigvee\limits_{j~\neq~i}p_j)~\rightarrow~\bigvee\limits^n_{i~=~0}
p_i \qquad (n~\geq~1)
\]

which are complete with respect to the class of all finite frames
of branching $n$. It is well known that they can be refuted in the
infinite comb. Nevertheless, not every frame containing the
infinite comb as a substructure refutes these axioms --- see
figure~\ref{r:bb2}. Therefore the following theorem is nontrivial:

\begin{figure}[b]
\begin{center}
\begin{picture} (90, 40)

    \put(0,0){\line(4,1){80}}
    \multiput(0,0)(20,5){5}{\circle*{1.5}}
    \multiput(85,22.5)(2.5,0.625){3}{\circle*{0.3}}

    \put(0,20){\line(4,1){80}}
    \multiput(0,20)(20,5){5}{\circle*{1.5}}

    \put(0,25){\line(4,1){60}}
    \multiput(0,25)(20,5){4}{\circle*{1.5}}

    \put(0,30){\line(4,1){40}}
    \multiput(0,30)(20,5){3}{\circle*{1.5}}

    \put(0,35){\line(4,1){20}}
    \multiput(0,35)(20,5){2}{\circle*{1.5}}

    \put(0,40){\circle*{1.5}}

    \put(0,40){\line(0,-1){20}}
    \put(20,40){\line(0,-1){15}}
    \put(40,40){\line(0,-1){10}}
    \put(60,40){\line(0,-1){5}}

    \multiput(0,15)(20,5){5}{\multiput(0,0)(0,-1.25){3}{\circle*{0.3}}}

\end{picture}
\caption{A structure containing an infinite comb as a substructure
where Gabbay-de Jongh axiom $\mbox{\boldmath$bb_2$}$ is true
\label{r:bb2}}
\end{center}
\end{figure}
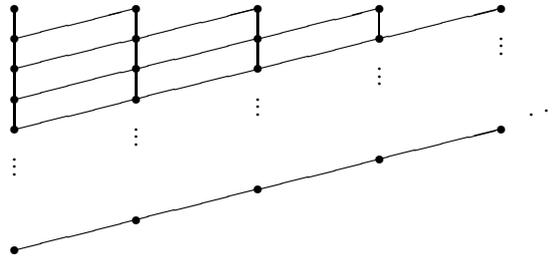

\begin{tm} \label{t:incompl}
If there exists a rule of the form
\begin{equation} \label{e:rul1}
\frac{\begin{array}{cccc}
(\psi~\vee~(\psi~\rightarrow~e(\chi)))~\rightarrow~ \chi \\
\psi~\leftrightarrow~\varsigma~\rightarrow~\tau \\
\varsigma~\vee~\tau~\rightarrow e(\varsigma)~\wedge~e(\tau) \\
\chi~\leftrightarrow~\psi~\vee~e(\tau)
\end{array}}{\chi}
\end{equation} which is not admissible in some intermediate logic $L$, then
in any class of frames adequate for $L$ (if there exists any)
there must exist a structure refuting $\mbox{\boldmath$bb_n$} (n
\geq 2)$. Thus, if $L$ contains any of Gabbay-de Jongh axioms, it
must be incomplete.
\end{tm}

{\sc Proof.} 
It may be carried out in a manner similar to that of
Shehtman~\cite{shehtman}, but it is needlessly complicated, e.g.
with a superfluous use of transfinite induction. Therefore I would
like to sketch a more elegant and intuitive proof. Assume then
that there is a frame ${\cal F}$ for $L$, a valuation ${\cal V}$
and a point $x$ in ${\cal F}$ such that $x~\nvDash_{\cal V}~\chi$.
It is easy to check that $x$ must be the root of the submodel of
$\langle {\cal F}, {\cal V}\rangle$ depicted by
picture~\ref{f:comb}. Now let me define a new valuation ${\cal B}$
based on ${\cal V}$ and inspired by figure~\ref{f:comb}:
\tmlei{Here is where the original 2002 text is edited.}
\textcolor{cobalt}{
\begin{eqnarray*}
{\cal B}(p_i)&:=&\bigcap\limits_{\forall m \in \omega\:n\,\neq\,3m + i}{\cal V}(e^n(\psi)).
\end{eqnarray*}
Axioms of $L$ and Figure~\ref{f:comb} ensure that sets ${\cal
B}(p_0)$, ${\cal B}(p_1)$ and ${\cal B}(p_2)$ are distinct and
non-empty. 
 It is easily seen that the consequent of
$\mbox{\boldmath$bb_2$}$ is refuted at $x$ under the valuation
${\cal B}$. Now suppose that there is some $y~\geqslant~x$ such
that some conjunct of the premise of $\mbox{\boldmath$bb_2$}$ is
\emph{classically} refuted at $y$, e.g.,
\begin{equation}  \label{eq:pimpl}
y~\vDash_{\cal B}~p_0~\rightarrow~(p_1~\vee~p_2)
\end{equation}
and
\begin{equation}  \label{eq:pjpd}
 y~\nvDash_{\cal B}~p_1~\vee~p_2.
\end{equation}
\refeq{pimpl} and \refeq{pjpd} taken together imply
\begin{equation}  \label{eq:pzpjpd}
 y~\nvDash_{\cal B}~p_0~\vee~p_1~\vee~p_2.
\end{equation}
We claim that
\begin{equation} \label{eq:chi}
\exists n \in \omega\: y~\nvDash_{\cal V} e^n(\chi). 
\end{equation}
To see this, assume \refeq{chi} does not hold, that is,  $e^n(\psi) \vee e^{n+1}(\tau)$ is $\cal{V}$-satisfied at $y$ for every $n$.  Pick the smallest $m$ s.t. $y~\nvDash_{\cal V}~e^m(\psi)$; it exists by \refeq{pjpd}.  This means $e^{m+1}(\tau)$ must be satisfied, thus yielding $y \vDash_{\cal V} e^{m'}(\psi)$  for every $m' > m$. As by the assumption on $m$ we have the same for every $m' < m$ as well, we thus contradict \refeq{pzpjpd}. \newline
Hence, we can pick the smallest $m$ s.t. $y~\nvDash_{\cal V}~e^m(\chi)$. Note that for any $m'~\leq~m$,  $y~\nvDash_{\cal V}~e^{m'}(\tau)$, and hence our assumption on $m$ holds only if for any $m' < m$, $y~\vDash_{\cal V}~e^m(\psi)$. We can find an infinite comb similar to the one in
Figure~\ref{f:comb}, but whose root this time is $y$ and  whose labelling is obtained by  replacing each formula in Figure~\ref{f:comb} by its suitably iterated $e$-substitution; think of the subframe generated by the $m$-th point up the trunk. It is consequently possible to find some (in fact, infinitely many) 
points from this comb classically refuting $p_0 \to (p_1~\vee~p_2)$, contradicting \refeq{pimpl}.} \hfill $\dashv$ \vspace{0.2
cm}

\tmlei{ The rest of the paper is left in the form it was written in 2002.}

It may be worth mentioning that rule~\ref{e:rul1} is as a matter
of fact inspired by the form  of axioms in Shehtman's later
paper~\cite{sh80}. In his paper from 1977~\cite{shehtman} the
axioms were more complicated and to make Shehtman's 1977 theorem a
consequence of theorem~\ref{t:incompl} --- as I am going to do ---
rule~\ref{e:rul1} should be replaced by the following one:
\begin{equation} \label{e:rul2}
\frac{\begin{array}{ccccc} (\psi~\vee~(\psi~\rightarrow~e(\chi)))~
\rightarrow~ \chi \\ \psi~\leftrightarrow~\varsigma~\rightarrow~
\tau \\ \tau~\rightarrow~e(\tau) \\
\chi~\leftrightarrow~\psi~\vee~ e(\psi)
\\ e(\psi)~\rightarrow~\psi~\vee~e(\tau)
\end{array}}{\chi}
\end{equation}

Now let me consider a family of formulas introduced by Shehtman:
\begin{eqnarray*}
\beta_{-1}&:=&p, \qquad \qquad \qquad \qquad \gamma_{-1}~:=~q, \\
\beta_{0}&:=&q~\rightarrow~p, \qquad \qquad \qquad
\gamma_{0}~:=~p~\rightarrow~q,
\\ \beta_{n+1}&:=&\gamma_{n}~\rightarrow~\beta_{n}~\vee~
\gamma_{n-1}, \\ \gamma_{n+1}&:=&\beta_{n}~\rightarrow~
\gamma_{n}~\vee~\beta_{n-1}, \\
\alpha_{n}&:=&\beta_{n+2}~\wedge~\gamma_{n+2}~\rightarrow
\beta_{n+1}~\vee~\gamma_{n+1} \qquad (n \in \omega), \\
\eta&:=&\alpha_{0}~\rightarrow~\alpha_{1}~\vee~\alpha_{2}, \qquad
\epsilon~:=~\alpha_{0}~\vee~\alpha_{1}, \\
\delta&:=&\eta~\rightarrow~\epsilon, \qquad \qquad \qquad
\kappa~:=~\alpha_{1} ~\rightarrow~\alpha_{0}~\vee~\beta_{2}.
\end{eqnarray*}

If $\varsigma$ stands for $\beta_2~\wedge~\gamma_2$, $\tau$ stands
for $\beta_1~\vee~\gamma_1$ and $e$ is defined as follows:

\[e(p) := q~\vee~(q~\rightarrow~p), \qquad e(q)~:=~p~\vee~
(p~\rightarrow~q),\]

then the following observation allows me to use a variant of
theorem~\ref{t:incompl} concerning rule~\ref{e:rul2} \\
\begin{tabular}[t]{lcl}
$\alpha_0$ & is of the form $\psi$, i.e. & $\varsigma~\rightarrow~
\tau$,
\\ $\epsilon$ & is of the form $\chi$, i.e. & $\psi~\vee~e(\psi)$,
\\ $\delta$ & is equivalent to & $(\psi~\vee~(\psi~\rightarrow~
e(\chi)))~\rightarrow~\chi$, \\ $\kappa$ & intuitionistically
implies & $e(\psi)~\rightarrow~\psi~\vee~e(\tau)$, \\ $\tau~
\rightarrow~e(\tau)$ & is an {\bf Int}-tautology.
\end{tabular} \vspace{0.2 cm}

Of course, it would also be possible to use
theorem~\ref{t:incompl} without any modification. In this case one
should define $\epsilon$ as $\alpha_0~\vee~\beta_2~\vee~\gamma_2$
or even $\alpha_0~\vee~\beta_2$, $\delta$ as
$(\alpha_0~\rightarrow~\alpha_1~\vee~\beta_3)~ \rightarrow~
\alpha_0~\vee~\beta_2$ and no $\kappa$ is needed at all.
Nevertheless, I am going to stick to the first paper of Shehtman
to make references easier; the paper from 1980~\cite{sh80} is less
known.

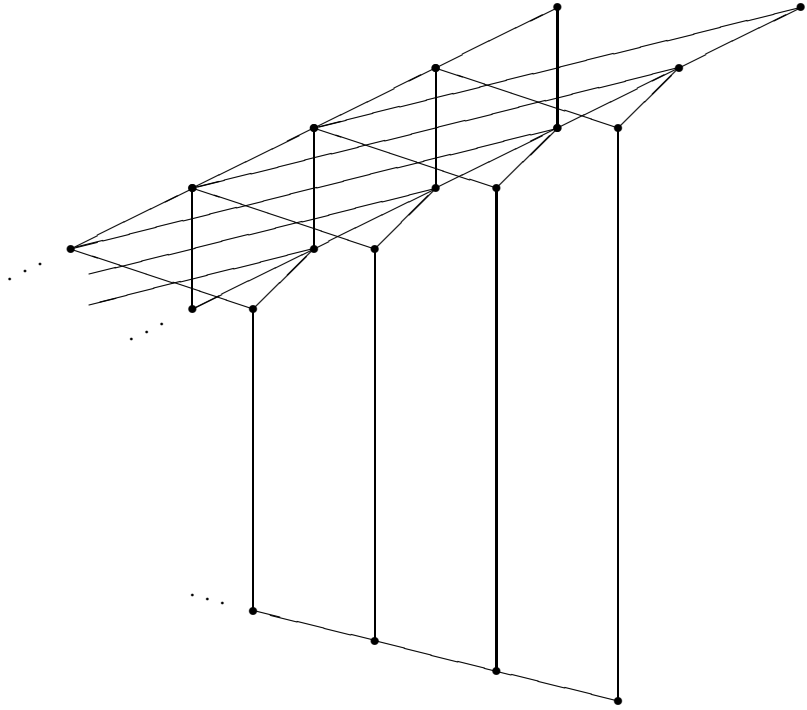
\begin{figure}
\begin{center}
\begin{picture} (110,120)

    \multiput(90,95)(-20,-10){3}{\usebox{\fsh}}
    \put(30,65){\usebox{\fshup}}

    \put(50,85){\line(-2,-1){40}}
    \put(50,85){\line(-4,-1){57}}
    \put(30,75){\line(-4,-1){37}}
    \multiput(-15,72.5)(-2.5,-1.25){3}{\circle*{0.3}}
    \multiput(5,62.5)(-2.5,-1.25){3}{\circle*{0.3}}

    \put(80,0){\line(-4,1){60}}
    \multiput(15,16.25)(-2.5,0.625){3}{\circle*{0.3}}

    \multiput(80,0)(-20,5){4}{\circle*{1.5}}

    \put(80,0){\line(0,1){95}}
    \put(60,5){\line(0,1){80}}
    \put(40,10){\line(0,1){65}}
    \put(20,15){\line(0,1){50}}

\end{picture}
\caption{The Fine frame  \label{r:fine}}
\end{center}
\end{figure}

\begin{lem} \label{t:sheh}
Axioms $\delta$ and $\kappa$ are true in a structure known as the
\emph{Fine frame} (see figure~\ref{r:fine}). Axiom
$\mbox{\boldmath$bb_n$}$ is true in a general frame based on the
Fine frame and generated by the two upward closed singletons. The
same general frame refutes axiom $\epsilon$.
\end{lem}

{\sc Proof.} It is quite easy and may be found, for example,
in~\cite{shehtman}. \hfill $\dashv$ \vspace{0.2 cm}

\begin{cor} [Shehtman]
An intermediate logic $L$ determined by axioms $\delta$,
$\kappa$, and $\mbox{\boldmath$bb_2$}$ is incomplete.
\end{cor}

{\sc Proof.} A consequence of theorem~\ref{t:incompl} and
lemma~\ref{t:sheh}. \hfill $\dashv$ \vspace{0.2 cm}

\begin{figure}
\begin{center}
\begin{picture} (125,60)

\multiput(0,0)(35,0){4}{\begin{picture}(20,30)
\multiput(10,0)(0,10){2}{\circle*{1}}
\multiput(0,20)(0,10){2}{\circle*{1}}
\multiput(20,20)(0,10){2}{\circle*{1}}
\multiput(0,20)(20,0){2}{\line(0,1){10}}
\put(10,10){\line(-1,1){10}} \put(10,10){\line(0,-1){10}}
\put(10,10){\line(1,1){10}}
\end{picture}}

\multiput(35,20)(35,0){3}{\usebox{\storey}}
\multiput(70,30)(35,0){2}{\usebox{\storey}}
\put(105,40){\usebox{\storey}}

\end{picture}
\caption{Frames ${\cal F}_{0}$, ${\cal F}_{1}$, ${\cal F}_{2}$,
${\cal F}_{3}$ \label{r:cont}}
\end{center}
\end{figure}
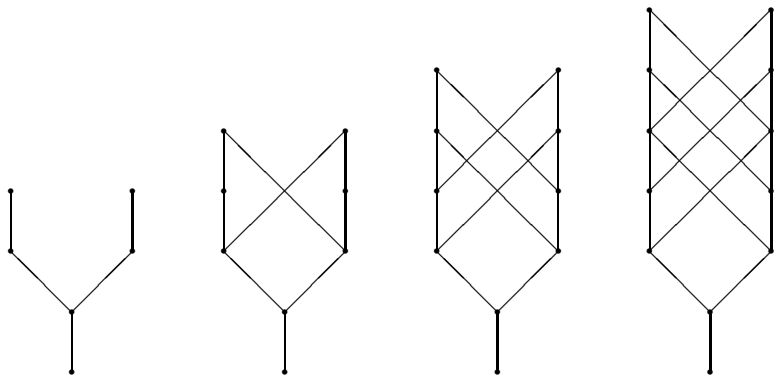

Now I may construct a continuum of incomplete logics inspired by
ideas from Kit Fine's classical
papers~\cite{fine74a},~\cite{fine74b}. I will construct a sequence
of frames ${\cal F}_n$ (see fig.~\ref{r:cont}) very similar to the
sequence from~\cite{fine74a}.

\begin{lem} \label{t:ax}
For any $n~\in~\omega$, ${\cal F}_n~\vDash~\delta~\wedge~\kappa~
\wedge~\mbox{\boldmath$bb_2$}$. Besides, ${\cal F}_n~\vDash~
\epsilon$.
\end{lem}

{\sc Proof.} The fact that the Gabbay-de Jongh axioms are true in
all of those frames is obvious. It is impossible to simultaneously
refute $\alpha_0$ and $\alpha_1$ in any of the frames, which
implies that ${\cal F}_n~\vDash~\delta~\wedge~\epsilon$. The
validity of $\kappa$ may be shown in the same way as in case of
the Fine frame. \hfill $\dashv$ \vspace{0.2 cm}

\begin{lem} \label{t:pm}
For any $n \in \omega$, there exists no p-morphism from any
generated subframe of ${\cal F}_n$ onto ${\cal F}_m~(m~\neq~n)$.
In other words, \[{\cal F}_n~\vDash~\beta^{\#}({\cal F}_m,~\bot)
(m~\neq~n),\] where $\beta^{\#}({\cal F}_m,~\bot)$ is a Jankov
formula for ${\cal F}_m$.
\end{lem}

{\sc Proof.} It is similar to the one in~\cite{fine74a} (by
induction). \hfill $\dashv$ \vspace{0.2 cm}

\begin{lem} \label{t:pmfine}
For any $n~\in~\omega$, there exists no p-morphism from any
generated subframe of the Fine frame onto ${\cal F}_n$. In other
words, Jankov formulas for the entire sequence are satisfied in
the Fine frame.
\end{lem}

{\sc Proof.} As above. \hfill $\dashv$ \vspace{0.2 cm}

\begin{tm} \label{t:main}
Distinct subsets of natural numbers generate distinct intermediate
logics whose axioms are $\delta$, $\kappa$,
$\mbox{\boldmath$bb_2$}$ and the Jankov formulas of those frames
from the sequence whose indices belong to a given subset of
$\omega$. All of these logics are incomplete.
\end{tm}

{\sc Proof.} The fact that these logics are all distinct is a
consequence of lemmas~\ref{t:ax} and ~\ref{t:pm}. The fact that
these logics are incomplete follows from theorems~\ref{t:incompl}
and lemmas~\ref{t:sheh} and ~\ref{t:pmfine}
--- a suitable inference rule is not admissible in any of the logics. \hfill
$\dashv$ \vspace{0.2 cm}

I would like to thank Professor A. Wro\'{n}ski, the supervisor of
my master's thesis, for his constant help and advice.

\end{document}